\documentclass[twocolumn,showpacs,preprintnumbers,amsmath,amssymb,prb,times]{revtex4}

\usepackage{graphicx}
\usepackage{pstricks}
\usepackage {ulem}

\begin{document}

\title{Magnetic response of nanoscale left-handed
metamaterials}

\author{R. S. Penciu$^{1}$, M. Kafesaki$^{1,2,*}$,
Th. Koschny$^{1,3}$, E. N. Economou$^{1,4}$, and
C.  M. Soukoulis$^{1,2,3}$}

\affiliation{$^1$Foundation for Research and Technology Hellas (FORTH),
Institute of Electronic Structure and Laser (IESL),
P.O. Box 1385, 71110 Heraklion, Crete, Greece}

\affiliation{$^2$Dept. Materials Science and Technology, University of
Crete, 71003 Heraklion, Greece}

\address{$^3$Ames Laboratory and Dept. Physics and Astronomy, Iowa State
University$^\dagger$, Ames, Iowa 50011, USA}

\address{$^4$Dept. of Physics, University of
Crete,  71003 Heraklion, Greece}

\email{kafesaki@iesl.forth.gr}

\begin{abstract}
Using detailed simulations we investigate the magnetic response of
metamaterials consisting of pairs of
parallel slabs or combinations of slabs with wires (including 
the fishnet design) as the length-scale of the structures is reduced
from mm to nm. We observe the expected saturation of the magnetic resonance
frequency when the structure length-scale  goes to the sub-micron
regime, as well as weakening of the effective
 permeability resonance and reduction of
the spectral width of the negative permeability region. 
All these results are explained by using an equivalent 
 resistor-inductor-capacitor
(RLC) circuit model, taking into account the current-connected kinetic
energy  of the electrons
inside the metallic parts through an equivalent inductance, added to the
magnetic field inductance in the unit-cell. Using this model 
we derive simple optimization rules for achieving  
optical negative permeability metamaterials of improved performance.
Finally, we analyze the magnetic response of the fishnet design and we
explain its superior performance
regarding the  high attainable magnetic resonance
frequency, as well as its inferior performance regarding the width
of the negative permeability region.

\end{abstract}
\pacs{ 41.20.Jb, 42.70.Qs, 81.05.Xj, 78.67.Pt}

\maketitle

\section{Introduction}

Left-handed metamaterials (LHMs), i.e. artificial composite structures with 
overlapping negative permittivity and permeability frequency bands giving 
rise to negative index of refraction 
\cite{Veselago68,SoukoulisAM}, have attracted 
recently an exponentially increasing attention. The main reason behind this 
attention is mainly the novel physical phenomena associated with those 
materials (negative refraction, opposite phase and energy velocity, reversed 
Doppler effect etc), which result to new capabilities in the manipulation of 
electromagnetic waves. Such an important capability is the superlensing 
capability of LHMs \cite{Pendry00}, i.e. the ability to offer 
subwavelength resolution imaging, which can have important implications in 
many scientific, technological and every-day life areas, like imaging, 
microscopy, lithography, ultra-compact data storage, etc.

Since the demonstration of the first left-handed material\cite{Smith00}, 
in 2000, 
operating in the microwave regime, many left-handed (LH)
structures have been 
created\cite{Aydin04,Shelby01,Katsarakis04,Ozbay08,Aydin08,Padilla06,Smith04}, 
and important efforts for the better 
understanding and the optimization of those structures have taken place. 
Among the various efforts within the LHM research, a large part has been 
devoted to the extension of the frequency of operation of LHMs from the 
microwaves to the optical regime, where the superlensing-based applications 
can find an important ground for manifestation. These efforts led to 
metamaterials with negative permeability operating in the few THz regime 
already in 2004\cite{Yen04,Katsarakis05}, which soon were followed 
by the first 
structures of 
negative permeability and/or negative index of refraction in the 
telecommunications regime and more recently in the lower visible regime. 
(For reviews of the existing research efforts on infrared (IR) and 
optical metamaterials see Refs. \cite{Soukoulis07a,Shalaev07}.)

While the first and most of the existing microwave LHMs are systems made of 
split-ring resonators\cite{Pendry99} 
(SRRs, i.e. interrupted metallic rings, giving rise to 
resonant loop-like currents, and thus to resonant permeability involving 
negative permeability values) and continuous wires (leading to the negative 
permittivity response\cite{Pendry96}), in most of todays high 
frequency LHMs the SRRs have 
been replaced by pairs of slabs (or stripes, or 
wires)\cite{Podolskiy03,Shalaev05,Dolling05,Zhou06b} - 
see Fig. 1(a). 
Like the SRR, the slab-pair also behaves as a resonant magnetic moment 
element, where the magnetic moment is created by resonant currents, 
antiparallel in the two slabs of the pair, forming a loop-like current. In 
most of the experimentally realized optical
slab-pair{\&}wire structures the slabs 
are as wide as the corresponding unit cell side and are physically connected 
with the wires, leading to a design known as 
fishnet\cite{Zhang05,Ulrich67,Kafesaki07,Helgert09} (see Fig. 1(d)). 
Fishnet design was able to give the highest in frequency LHMs up to 
now\cite{Chettiar07,Dolling07}.

The main reason behind the replacement of SRR by the slab-pair for the high 
frequency metamaterials, apart of slab-pair simplicity in fabrication 
(which is 
also a crucial parameter), is its ability to exhibit negative permeability 
response for incidence normal to the plain of the pair; this makes 
possible the 
demonstration of the negative permeability response with just a monolayer of 
slab-pairs. Indeed, up to now most of the demonstrated ``magnetic'' 
metamaterials (i.e. materials of resonant and negative permeability) and 
LHMs are single layers, while only few multilayer samples have been 
fabricated\cite{Katsarakis05,Liu09}. (Note that in the optical 
regime what is difficult to be 
achieved is the negative permeability component of a LHM, since the negative 
permittivity response can be easily obtained using metals; that is why most 
of the existing efforts to go to the optical LHMs start from attempts to 
achieve structures of only negative permeability.)

Since many of the existing attempts to create high frequency magnetic 
metamaterials and LHMs are based on the scaling down of known 
microwave designs, there are already  efforts trying to 
determine the possibilities and the limitations of this scaling 
approach\cite{Zhou06a,Tretyakov07,Ishikawa07,Sarychev06}. 
(Note that the properties of the metals, which are involved in most of 
todays metamaterials, are drastically different in the optical regime 
compared to microwaves - there, metals behave almost as perfect conductors.) 
Most of those attempts concern SRR 
systems and they have led 
to two important conclusions\cite{Zhou06a,Economou09,Soukoulis07b}:

(a) By scaling down a SRR, the frequency of its resonant magnetic 
response does not continuously increase, but after some length scale it 
saturates to a constant value. This value was found to be dependent on the 
SRR geometry employed, and with proper modifications of this geometry 
(e.g., adding gaps in the SRR) it could go up to a small fraction 
(e.g. 20$\%$) of the plasma frequency, i.e. to the middle visible range. The 
saturation response of the magnetic resonance frequency was explained taking 
into account the contribution of the kinetic energy of the electrons 
associated with the current inside 
the SRR ring to the magnetic energy created by the loop current (or, 
equivalently, taking into account the dispersive response of the metal in 
the conductivity). 

(b) The magnetic permeability resonance becomes weaker and weaker by going to 
smaller length-scale SRR systems, and below some length scale it 
ceases to reach negative values. This weakening of the 
permeability was attributed to the kinetic energy of the electrons (giving
rise to saturation) in combination with the  
increased resistive losses in the metal as one goes to higher 
frequencies; these losses are strengthened by the resonant response, implying 
long-time interaction of the wave with the metallic structures.

Although the existing studies are very revealing concerning the high 
frequency response of magnetic metamaterials, they examine the influence of 
the kinetic energy (or the dispersive response of the metals) only to the 
magnetic resonance frequency and not to other features of the resonant 
magnetic response (like resonance shape and damping factor), neither 
clarified the role of the losses in the saturation of the magnetic resonance 
frequency. Moreover, the role of the 
various geometrical parameters in the high frequency response of 
metamaterials still remains to be determined, as to identify the 
dominant parameters determining this 
response and to define optimization rules for those materials.

\begin{figure}[htbp]
\includegraphics[width=3.41in]{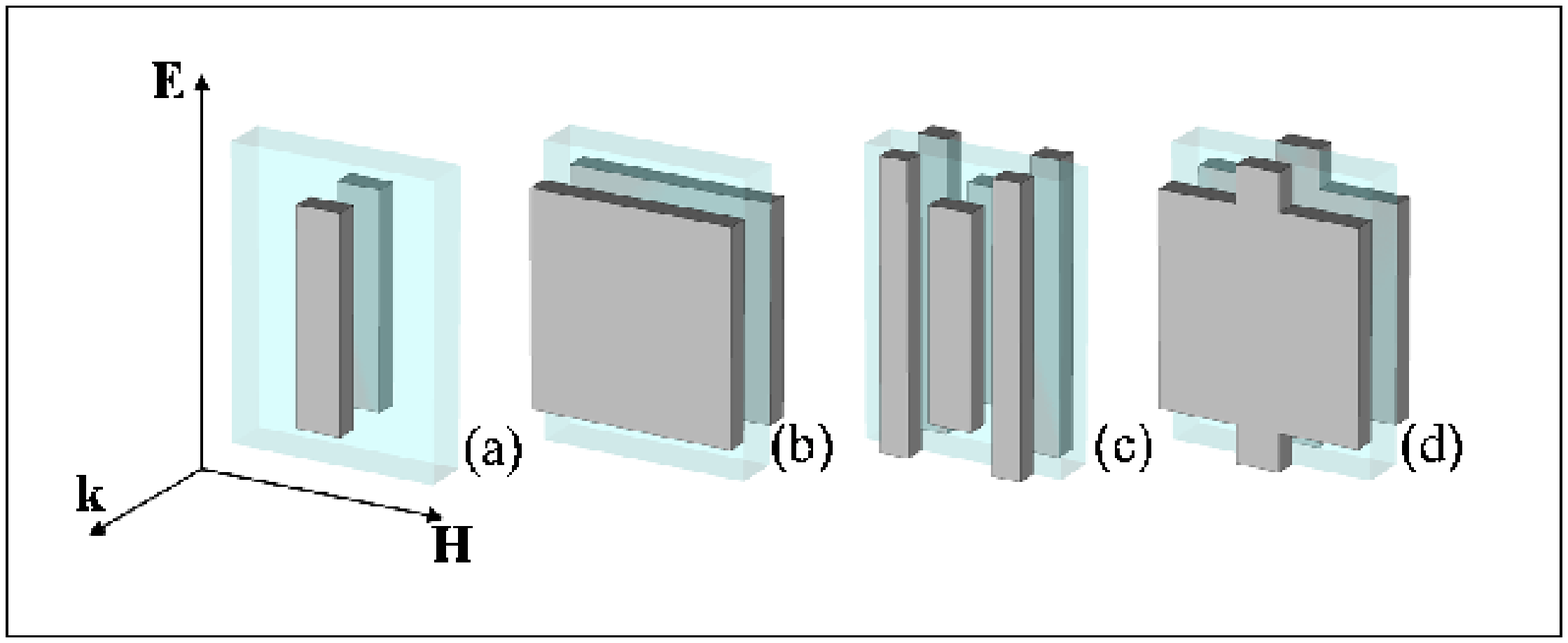}  
\includegraphics[width=2.0in]{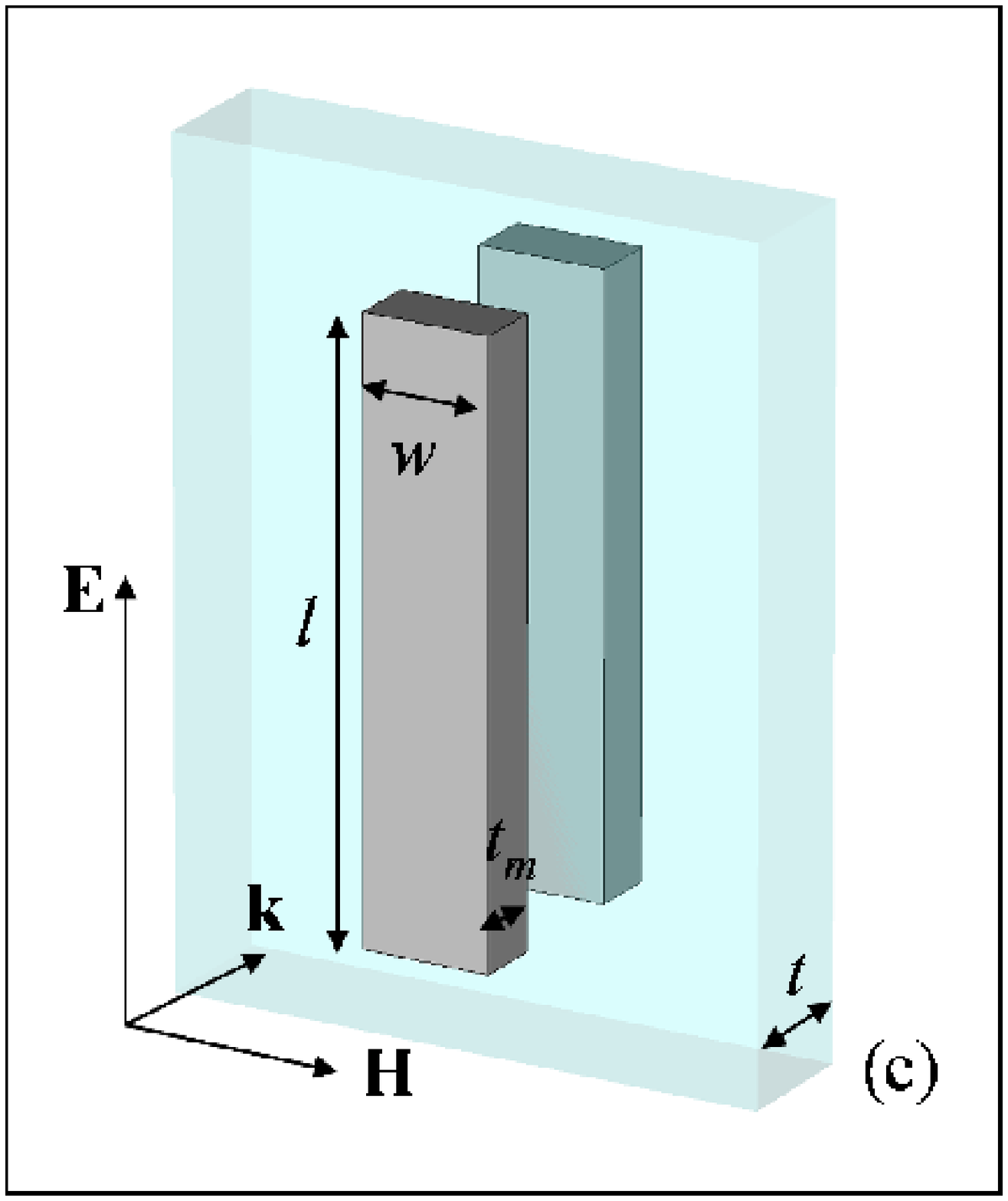}  
\caption{The unit cell of the 
four designs studied. (a): Narrow-slab-pair system; (b) wide-slab-pair 
system; (c) slabs{\&}wires system; (d) fishnet design. Panel 
(e) is a magnification of panel (a) where the structure parameters 
appearing in the structure simulations are shown. The parameters of 
the pair are given as a function of the scale parameter $a=a_{\bf k}$ 
(lattice constant along propagation direction; equal to the system 
thickness): lattice constants $a_{\bf E}=2.97 a_{\bf k}$, 
$a_{\bf H}=2.19 a_{\bf k}$, 
slab-length $l=2.19a_{\bf k}$, 
slab-width $w=0.47 a_{\bf k}$, 
thickness of the metal $t_{m}=0.25 a_{\bf k}$ and thickness of 
the substrate $t=0.5 a_{\bf k}$. For the fishnet design (d) the width of
the slabs is equal to the corresponding unit cell side ($a_{\bf H}$), while
the width of the ``necks'' (continuous metallic parts joining the slabs
along ${\bf E}$ direction) is $w_n=0.469 a_{\bf k}$.
 The dielectric spacer separating the metallic
pairs has been considered as glass (with relative permittivity 
$\varepsilon_b=2.14$), while for the metal plasma 
frequency and damping factor the aluminum parameters have been employed.} 
\label{fig1}
\end{figure}

In this paper we attempt to study all the above mentioned issues. We will be 
restricted to systems based on pairs of slabs, alone or in combination with 
continuous wires; this is mainly due to the fact that slab-pair-based 
systems are 
offered for an easy experimental demonstration of negative permeability or 
negative index response and, moreover, have been proven up to now
 the most promising 
systems for the achievement of high frequency negative permeability and 
negative index metamaterials. The structures 
discussed here are shown in Fig. 1. Following the approach of 
Refs. \cite{Zhou06a,Soukoulis07b}, we 
will attempt to analyze the high frequency magnetic response of those 
structures, to compare their performance and to propose optimization 
rules for them. For 
that we examine in detail the scaling behavior of the magnetic resonance 
frequency and the magnetic permeability as the structures are scaled down 
from mm to nm scale.

The basic idea that we will use to reproduce and understand the small length 
scale (high frequency) behavior of our structures is the consideration of 
the kinetic energy of the current carried electrons. This kinetic energy, as 
being proportional to the square of the velocity (and thus of the frequency, 
just like the magnetic energy), is added to the magnetic energy of the 
structures and in small length-scales it dominates the magnetic 
metamaterials response. The consideration of this kinetic energy is done 
here through an equivalent ``kinetic'' inductance\cite{Solymar,Zhou06a} 
(or electrons' inductance), 
added to the magnetic 
field inductance in an effective resistor-inductor-capacitor (RLC) 
description of the artificial magnetic structures.

Specifically, the paper is organized as follows: In Section II we 
present the ``high frequency'' response of our structures, as revealed by 
numerical simulations concerning the magnetic resonance frequency, the 
form of the magnetic permeability resonance and the losses. The wave 
propagation 
characteristics in those structures are analyzed and explained in Section 
III, using an effective RLC description of the structures and taking 
into account the dispersive behavior of the metals through the kinetic metal 
inductance. Based on the results and analysis of Sections II and III, 
in Section IV we present 
basic optimization rules for the achievement of high frequency magnetic 
metamaterials and left-handed materials with improved performance. 
There we discuss also the  fishnet design, which has been
proven up to now the optimum design for achievement of optical negative
index materials.  
Finally, in Section V we show that the simple RLC circuit model
does not only have qualitative power, but it can be used also to give
quantitative results if plugged with accurate relations for the capacitance
and the inductance of the system. 

\section{Numerical simulations}

In this section we present calculation results concerning the scaling of the 
magnetic resonance frequency and the magnetic permeability of the structures 
shown in Fig. 1. The geometrical parameters used in the simulations
 are those 
mentioned in Fig. 1;  for the permittivity of 
the metal the Drude dispersion model has been employed, i.e. 
$\varepsilon=\varepsilon_0 
[\varepsilon^{(0)}-\omega_{p}^2/(\omega^2+i\omega \gamma_m)]$, 
with the 
parameters of the aluminum (plasma frequency $\omega_p=22.43\times 10^{15}$ 
sec$^{-1}$,
collision frequency $\gamma_m=12.18\times 10^{13}$ sec$^{-1}$), and 
$\varepsilon^{(0)}=1$.
Using Drude dispersion model one takes automatically into account the 
mass and any kinetic
energy of the current-carrying electrons inside the metal
(the contribution of the bound electrons is also taken into account,
through the constant $\varepsilon^{(0)}$).

The calculations presented here have been performed using the Finite 
Integration Technique, employed through the MicroWave Studio (MWS) 
commercial software. Using MWS, the transmission and reflection from a 
monolayer of the structure have been obtained; these data have been used for 
the determination of the effective permittivity and permeability of the 
structures, through a standard retrieval procedure based on a homogeneous
effective medium approach\cite{Smith02,Smith05}. 

\begin{figure}
\label{fig2}
\centerline{\includegraphics[width=3.20in]{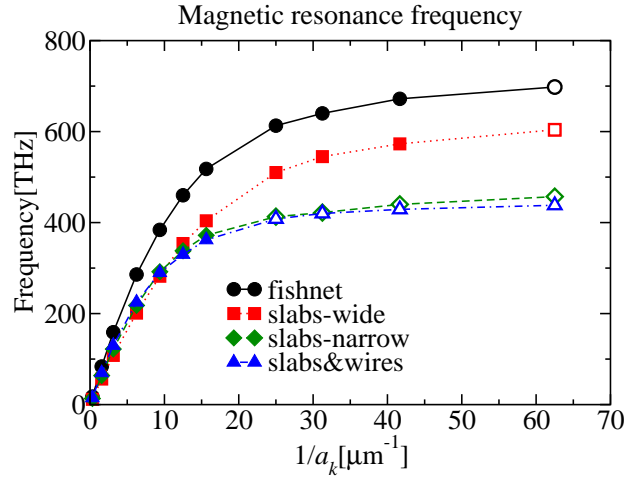}} 
\caption{Scaling of the magnetic resonance frequency with the linear 
size of the unit cell along propagation direction ($a_k$) for the four designs
of Fig. 1: Fishnet 
(black circles), wide-slabs (red squares), narrow-slabs (green diamonds), 
slabs\&wires (blue triangles). The solid symbols indicate the
existence of negative permittivity values, while the open symbols indicate
that the permeability resonance is weak and unable to reach
negative values for the Re($\mu$).}
\end{figure}

\begin{figure}
\label{fig3}
\includegraphics[width=2.50in]{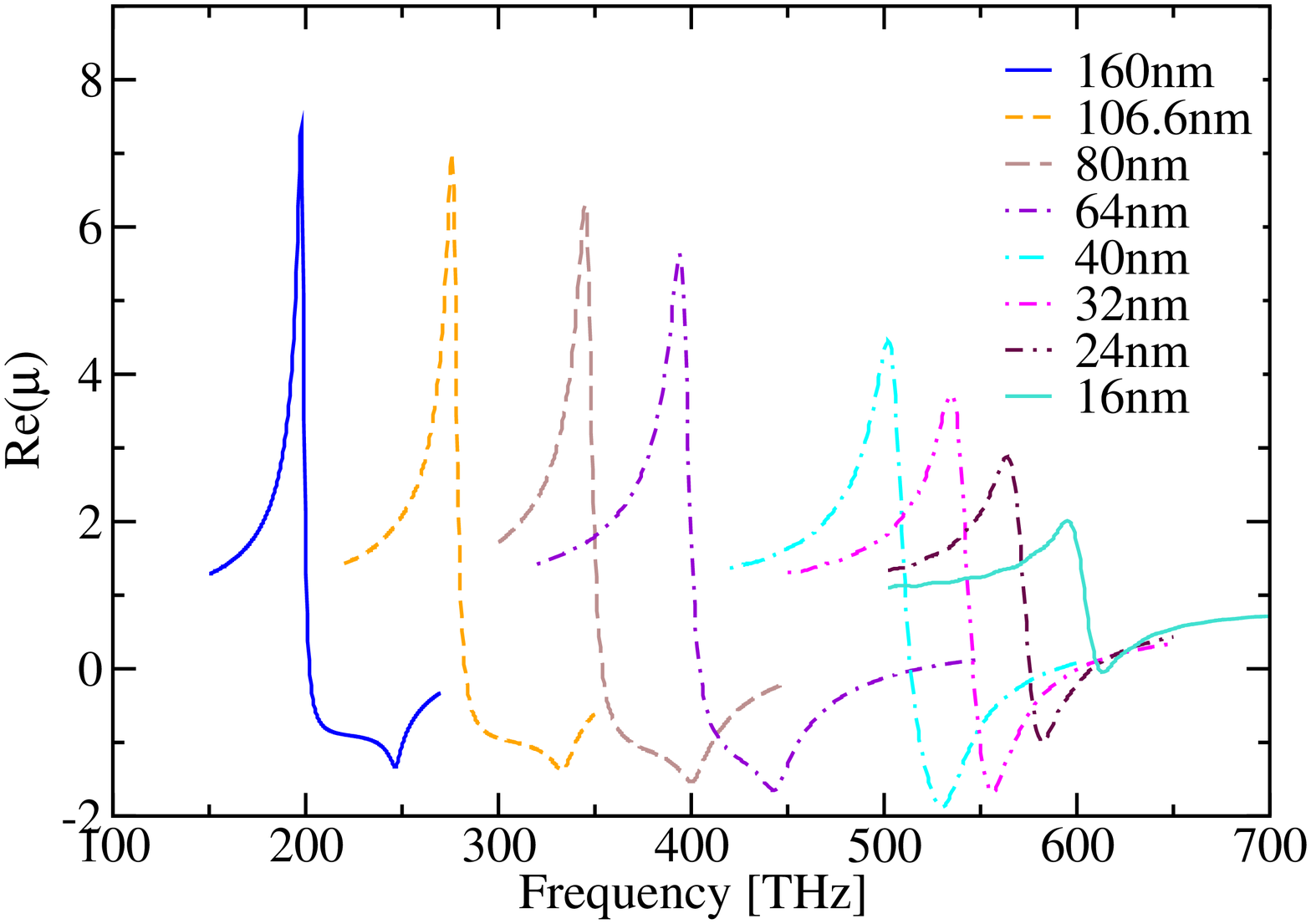} \\
\includegraphics[width=2.50in]{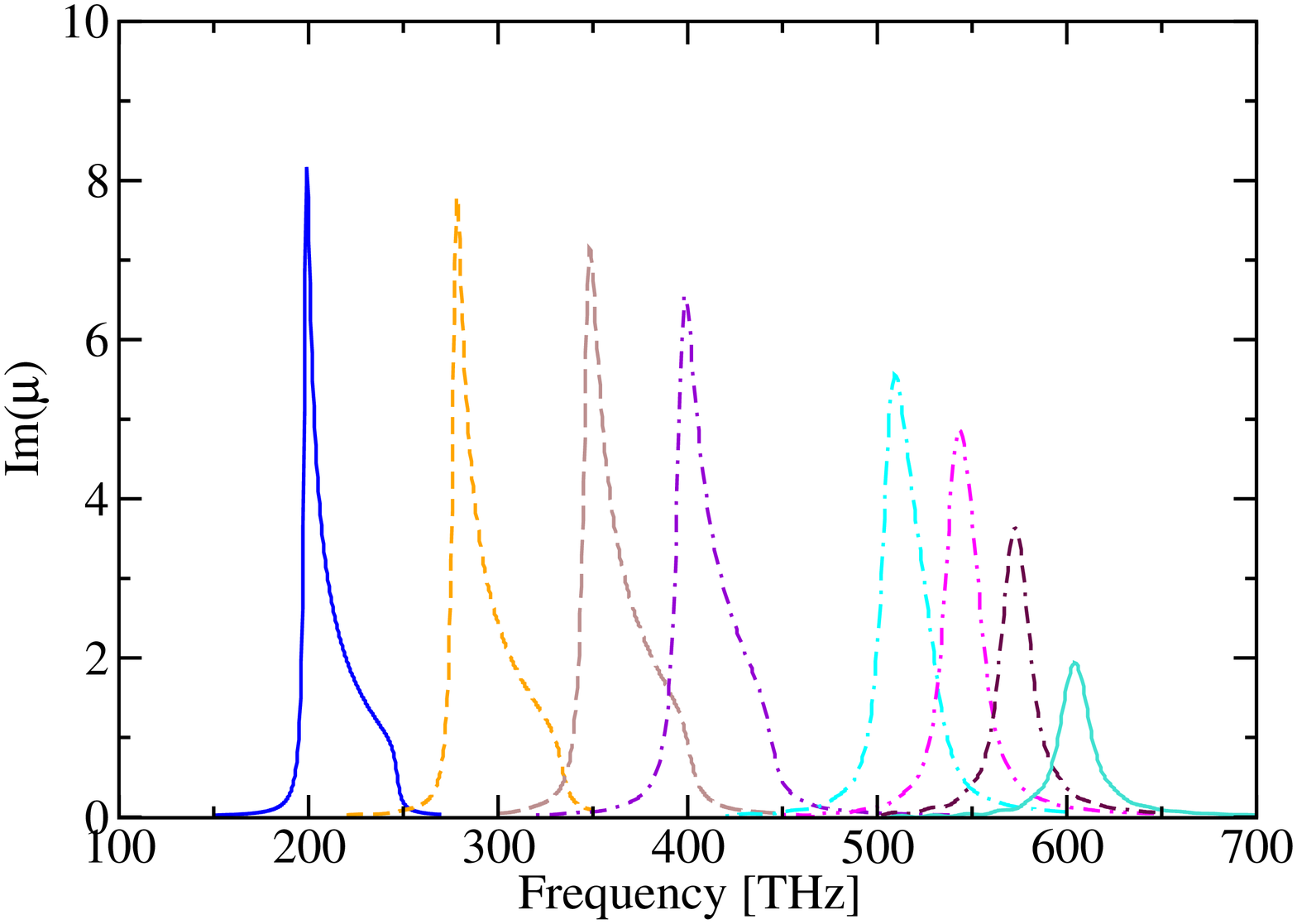} 
\caption{Frequency dependence of the real (top panel) and the
imaginary (bottom panel) part of the 
magnetic permeability of the wide slab-pair system for various length scales.
The legends indicate the lattice constant along propagation direction
for each specific system.
}
\end{figure}

In Fig. 2 we present the magnetic resonance frequency for the four 
structures of Fig. 1 as the structures are scaled down from mm to nm scale; 
in Fig. 3 we show the real and imaginary part of the magnetic permeability 
as a function of frequency, for wide-slab-pairs (structure
of Fig.~1(b)) of various length scales.

As can be seen in Fig. 2, in slab-pair systems we observe the same behavior 
as the one reported earlier for SRR structures\cite{Zhou06a,Soukoulis07b}: 
while in larger length-scales the 
magnetic resonance frequency scales inversely proportional to the structure
linear size, at frequencies in the near infrared (IR) towards optical 
regime this linear
scaling breaks down, and the magnetic resonance 
frequency saturates to a constant value. This saturation value is different 
for the different designs employed, with larger the one of the fishnet 
structure. Note  that in the slabs{\&}wires case (structure of Fig. 
1(c)) the presence of wires does not affect the magnetic resonance frequency 
of the slabs, while in the fishnet structure the presence of wires leads to 
higher saturation value for the magnetic resonance frequency. This behavior of
the fishnet design will be discussed and explained in Section IV.

It is important to mention here that the saturation values for the magnetic 
resonance frequency of the slab-pair-systems are in all cases larger than the 
saturation values obtained for SRRs of a single gap\cite{Klein06} 
(like, e.g., U-shaped SRRs; single-gap SRRs are  the 
only SRR-based system that has been fabricated in the nm scale), indicating
once more the suitability of the slab-pair-based systems for the
achievement of optical magnetic metamaterials.

Concerning the permeability results shown in Fig.~3, we observe that, just 
like the SRRs\cite{Zhou06a,Soukoulis07b}, as the length scale of the 
structure becomes smaller 
 the permeability resonance becomes weaker, ceasing ultimately to 
reach negative values. This weakening is revealed in both the 
real and imaginary part of the permeability resonance, and it will
be analyzed in more detail and quantified  in the following
paragraphs. (Note that the ``truncation'' of the resonances of Fig. 3, 
which is observed in  the larger scale structures, is a result of the 
larger influence of the periodicity in these length scales\cite{Koschny05}. 
In smaller scales this influence becomes smaller, due to the deeper
sub-wavelength scale of the corresponding structures and the weaker 
resonant response, which results to smaller
effective index and thus larger wavelength inside the structures.)

One quantity which is of great interest in left-handed materials and it
is strongly affected by the weakening of the permeability resonance is the
width of the negative permeability regime, which  roughly corresponds
with the operational band width of a LHM.
In Fig. 4 we present the relative band-width (i.e. band-width divided 
by the lower frequency of the negative 
permeability band)
for the four
structures shown in Fig. 1.  As can be seen in Fig. 4,
the operational band-width, which in larger scales is almost 
independent of the length scale, in smaller scales it is strongly
reduced,  ultimately going to zero for all 
designs. This shows that the
negative permeability is ultimately killed in the nm scale structures. 
Among our four structures the wide-slab-pair one is characterized
by the larger band-width. Finally, it is 
 worth-noticing the reduced band-width of the  fishnet design, compared
to the band-width of the slab-pair-only structure.
This bandwidth behavior will be discussed in the next section.

Since a main issue for the achievement of high frequency metamaterials of 
satisfactory performance is the losses, due to the increased resistive 
losses in the metallic components going to the optical regime, we performed a 
detailed analysis of the losses in high frequency metamaterials, trying to 
estimate which aspects of the high frequency metamaterial response are 
mainly affected by the resistive losses, and, consequently, to seek  ways 
to minimize the influence of those losses, using proper design 
modifications.

As a first step we calculate the losses as a function of frequency for
various length scales of our systems. In Fig. 5 we present these losses 
for the wide-slab-pair system of Fig. 1(b).
 The losses, $A$, have been calculated through the 
relation $A=1-R-T$, where $R$ and $T$ are the reflection and 
transmission coefficient, 
respectively, through one unit cell of the structure along propagation 
direction. As is expected, the losses show a dramatic increase by going to 
smaller length scales and higher frequencies. This increase seems to have
an exponential dependence on the  magnetic resonance frequency.
\vspace{2mm}

\begin{figure}
\label{fig4}
\centerline{\includegraphics[width=3.00in]{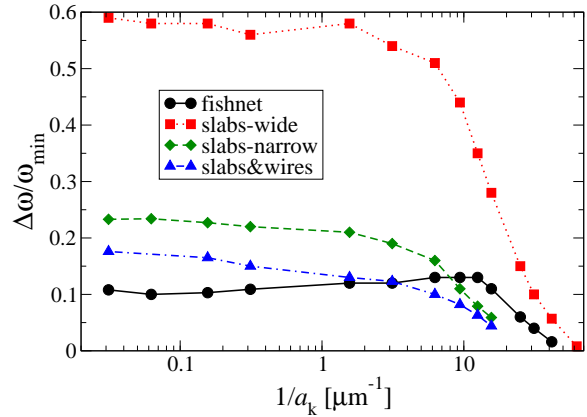}} 
\caption{Band-width, $\Delta \omega$, of the negative magnetic 
permeability regime vs the inverse unit cell size, $1/a_{\bf k}$, for 
fishnet  (black circles), wide-slab-pairs (red squares),   
narrow-slab-pairs (green diamonds) and slabs{\&}wires (blue triangles). 
The band-width is normalized by the 
minimum frequency (lower limit) of the negative permeability 
regime, $\omega_{\rm min}$. $a_{\bf k}$ is the lattice constant along propagation
direction.}
\end{figure}

\begin{figure}
\label{fig5}
\centerline{\includegraphics[width=3.20in]{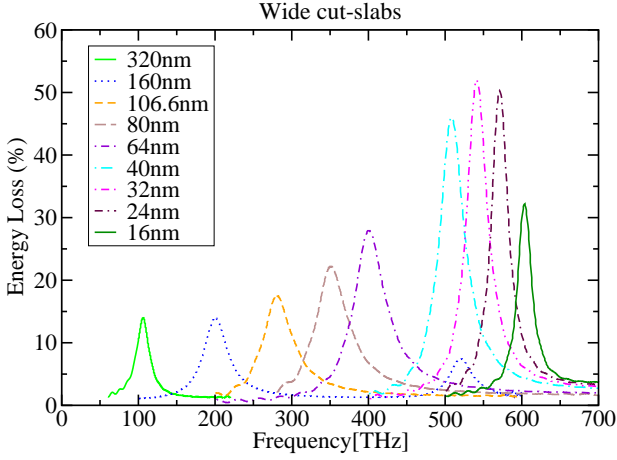}} 
\caption{Loss, $A$, per 
unit cell ($\%$) for the wide-slab-pair system for various 
length scales of the system, close to the magnetic resonance 
frequency saturation regime. The legends indicate the  
lattice constant along propagation
direction.}
\end{figure}

From Figs 2-5, one can see clearly the decreased performance of our 
structures going to nm scale, inhibiting their ability to give high quality 
optical left-handed 
metamaterials. An interesting question arising here is the role 
of the resistive losses on this decreased performance.
To examine this role, we repeated the above 
shown calculations  using for the 
metal a $2\pi \times 1000$ times reduced collision frequency 
compared to that of 
aluminum, i.e. $\gamma_m=12.18 \times 10^{10}/2\pi$. The results 
concerning the 
wide-slab-pair system are shown in Fig. 6.

Specifically, Fig. 6(a) shows the saturation of the magnetic resonance 
frequency and Fig. 6(b)
the real part of the resonant permeability response for 
the smaller length scales.
Comparing the result of Fig. 6 with those for the non-reduced value
of $\gamma_m$, one can 
see that the saturation of the magnetic resonance frequency seems to be 
totally unaffected by the value of the metal collision frequency. 
This indicates 
that the loss-factor of the metal employed is not able to affect the highest
achievable magnetic resonance frequency of each specific design. On the 
other hand, the metal loss-factor can affect the minimum length scale able 
to give negative permeability response, as shown comparing Fig. 6(b) with 
Fig. 3(a). Indeed, in the small $\gamma_m$ cases the magnetic 
permeability resonance is 
quite stronger, maintaining negative values up to smaller length scales. 
Although the strength of the resonance (as measured, e.g., 
by the minimum value of the Re($\mu$)) seems to be strongly affected by the 
$\gamma_m$ value, calculating the width of the negative permeability 
regime (in the cases that such a regime exists), $\Delta \omega$ - not shown
here -, it is observed that it is  only slightly affected by
the $\gamma_m$ value; 
it tends to zero for both high and low values of $\gamma_m$,
indicating that  even in the absence of resistive losses one can 
not go to arbitrarily high-frequency negative permeability response 
over a practical band-width.

\begin{figure}
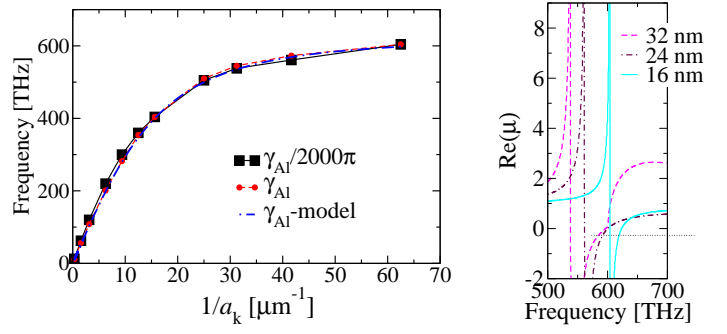

\centerline{\includegraphics[width=2.30in]{Fig6a.eps} \hspace*{4mm}
\includegraphics[width=1.1in]{Fig6b.eps}
}
\caption{
(a): Saturation of the magnetic resonance frequency for the wide-slab-pair 
structure using for the metal a 2000$\pi$ times reduced 
collision frequency compared with that of aluminum, 
i.e. $\gamma_m=\gamma_{Al}/2000\pi$ (solid-black line with squares). 
For comparison the corresponding result
for $\gamma_m=\gamma_{Al}$ is given (dashed-red line with circles).
Dotted-dashed line (blue color) shows the scaling of the magnetic resonance
frequency as it is obtained from our RLC circuit model (for the 
$\gamma_{Al}$ case).
(b): Magnetic permeability resonance (shown in ${\rm Re}(\mu)$) 
as a function of frequency for the wide-slab-pair structure of Fig. 1(b),
for  $\gamma_m=\gamma_{Al}/2000\pi$
 and for three different structure length-scales. Legends denote the lattice
constant along propagation direction, $a_{\bf k}$.}
\end{figure}

The results presented above raise many questions concerning the high-frequency 
magnetic metamaterial response and the main phenomena and factors 
determining this response. Before attempting an interpretation of this 
response we will summarize here the main effects observed so far and the 
main questions that one needs to address as to clarify and to
be able to predict 
the existence or performance of optical negative magnetic 
permeability response:

(a) The magnetic resonance frequency of slab-pair-based systems, while in mm 
scale structures scales inversely proportional to the structure length-scale, 
in sub-$\mu$m 
scale structures saturates to a constant value, depending mainly on the 
geometry of the structure and independent of the resistive losses in the 
component materials. Among the structures that have been studied here the 
fishnet design leads to the higher saturation frequency. Moreover, while in 
the slabs{\&}wires design the magnetic resonance frequency is almost 
unaffected by the presence of the wires, it is not the same in the fishnet 
design, where the saturation value is quite higher than the saturation value 
for the the slabs  only case.

(b) The magnetic permeability resonance in sub-$\mu$m scale
structures becomes 
more and more weak by reducing the structure length-scale, and ultimately 
ceases to reach negative values. The length scale at which $\mu <0$ stops to 
exist depends on the structure geometry (design) and on the resistive losses 
in the component materials, especially in the metallic parts. Among the 
structures that we have examined, negative permeability in smaller length
scales is 
achieved in the slab-pair structures, not associated though with higher 
magnetic resonance frequency. Negative permeability at higher frequencies is 
achieved in the fishnet design.

(c) Concerning the relative spectral width of the negative 
permeability regime, 
this width, while it is constant 
in mm scale structures, as 
one goes towards nm scales it becomes smaller and 
smaller, up to almost vanishing. This 
width depends on the structure geometry  and seems only slightly  
dependent on the 
resistive metallic losses.

\section{Analytical model and interpretation of the results }


To understand and explain the results presented in the previous section, we 
use the common approach of describing the artificial magnetic 
structures (at the resonant magnetic response regime)
as equivalent effective RLC 
circuits. Using circuit theory and basic 
electromagnetic considerations\cite{SoukoulisAM}, one can easily 
obtain an expression for 
the frequency dependence of the effective magnetic permeability, 
$\mu (\omega)$, for a lattice of ``magnetic'' elements\cite{SoukoulisAM}.

For the explanation of the high frequency magnetic response 
we follow the approach of Ref.\cite{Zhou06a}, based on the 
consideration of the kinetic 
energy of the electrons in the metal, $E_k$, besides the magnetic 
energy of the 
resulting electromagnetic field. Thus, we replace the magnetic inductance 
of the 
system in the effective RLC circuit description by the total inductance 
resulting as a 
sum of the magnetic field inductance, $L$, and the kinetic inductance, 
$L_{e}$, where $L_e$ is
defined by $E_k = N_e m_e v_e^2 / 2 = L_e I^2 / 2$ ($N_e$ is the
total number of electrons, $m_e$ is the electron mass and $v_e$ the average
electron velocity).

Assuming a slab-pair system like the one shown in Fig. 1(e), of length 
$l$ and slabs
separation $t$, excited by a magnetic field of the form $H = H_0 e^{ - i\omega 
t}$and direction as shown in Fig. 1(e), and applying the Kirchhoff voltage 
rule, one can obtain

\begin{equation}
\label{eq1}
(L + L_e )\ddot {I} + \frac{1}{C}I + \dot {I}R = - \ddot {\phi } = 
\omega^2 \mu _0 lt H_0 e^{ - i\omega t}.
\end{equation}
$\phi $ is the external magnetic flux, 
$\phi = \mu _0 lt H$, and $R$ and $C$ the resistance 
(frequency independent, accounting for the ohmic losses) and the 
capacitance of the system, respectively. The obvious 
solution of Eq. (\ref{eq1}) is $I=I_0{\rm e}^{-i\omega t}$, with

\begin{equation}
\label{eq2}
I_0=-\frac{\omega^2[\mu_0 l t/(L+L_e)]}{\omega^2-1/(L+L_e)C+i\omega R/(L+L_e)} 
H_0.
\end{equation}

Having the current one can easily obtain the pair magnetic dipole moment, 
$m={\rm area}\times{\rm current}=ltI$, and the magnetization $M=(N_{LC} 
/V)ltI=(1/V_{uc})ltI$, $N_{LC}$ is the number of ``RLC'' circuits 
in the volume $V$, and 
$V_{uc}=a_{\rm {\bf E}} a_{\rm {\bf H}} a_{\rm {\bf k}}$ 
is the volume per unit cell, where $a_{\rm {\bf E}}, a_{\rm {\bf H}}, 
a_{\rm {\bf k}}$ are the system lattice constants along the
{\bf E}, {\bf H}, and {\bf k} direction respectively 
($a_{\rm {\bf H}} \ge w,\,a_{\rm {\bf E}} > l,\,a_{\rm {\bf k}} \ge t)$ 
- see Fig. 1). 

Finally, using $M=\chi_m(\omega)H, \mu(\omega)/\mu_0=1+\chi_m(\omega)$, 
with $\chi_{m}$ the magnetic susceptibility, one obtains that 

\begin{equation}
\label{eq3}
\mu(\omega)=\mu_0 [1-\frac{(1/V_{uc})(\mu_0(lt)^2/(L+L_e))\omega^2}{\omega^2
-\omega_{LC}^2+i\omega\gamma}], 
\end{equation}
with 
\begin{equation}
\label{eq3b}
\omega_{LC}=\frac{1}{\sqrt{(L+L_e )C}}, \hspace{4mm}
\gamma=\frac{R}{L+L_e}. 
\end{equation}

\noindent
$\omega_{LC}$ is the magnetic resonance frequency of the system 
and $\gamma$ is the 
dumping factor, representing all the losses and the scattering mechanisms.

The inductance $L_e$ can be easily calculated by calculating the kinetic 
energy of the electrons, 
$E_k=N_e m_e v_e^2 /2=V_w n_e m_e v_e^2 /2$, 
and expressing the velocity through the current, 
$I=e w t_{m} n_{e}v_{e}$ ($n_{e}$ is 
the number density of free electrons, $e$ is the electron charge, and
$V_{w}=wt_{m}l$ is the volume of the metallic slab - see Fig. 1(e)). 
This way, one can obtain
$L_e =lm_e /wt_m e^2n_e =(l/wt_m )(1/\omega_p^2 \varepsilon_0)$, 
where $\omega_p=\sqrt{e^2 n_e/m_e \varepsilon_0}$ is the plasma frequency of 
the bulk metal.

Note that  exactly the same results as in Eq. (\ref{eq3}) can be
obtained by considering, instead of the kinetic energy for the
derivation of $L_e$, the dispersive behavior of the metal conductivity. 
Indeed, starting with the frequency dependent 
Drude-type conductivity 
\begin{equation}
\label{eq4}
\sigma =i\varepsilon _0 \frac{\omega _p^2 }{\omega+i\gamma_m },
\end{equation}
we obtain for the 
total resistance 
\begin{equation}
\label{eq5}
R_{tot} =\frac{1}{\sigma }\frac{l}{S}=(\frac{\gamma_m }{\varepsilon_0 \omega_p^2}-i\frac{\omega}{\varepsilon_0\omega_p^2})\frac{l}{S}=R-i\omega L_e.
\end{equation}
Using $R_{tot}=R-i\omega L_{e}$ in Eq. (\ref{eq1}) in the place of 
$R$ and only the 
magnetic field inductance (in order to avoid counting twice the kinetic
inductance $L_e$), one can obtain the same current solution as in 
Eq. (2).

In the following, to simplify our discussion we will consider the magnetic 
field inductance to be given by the inductance of a solenoid of 
area $lt$ and 
length $w$ (see Fig. 1), 
i.e. $L=\mu_0 lt/w$, and the capacitance by that of a parallel 
plate capacitor of area $wl/2$, plate separation $t$ and dielectric core
of relative dielectric constant $\varepsilon_b$,
i.e. $C=\varepsilon _0 
\varepsilon_b (wl)/t$. (Note that these formulas are appropriate for the 
case of wide slabs while describe much less satisfactory the narrow 
slabs case; in any case there is a numerical correction
factor of the order of one - see also Section V.
Note also that in our discussion we will not consider any inter-unit-cell 
capacitance\cite{Zhou06b}, which is important in the case of quite long slabs, 
i.e. slabs that  approach the unit cell 
boundaries along \textbf{E} direction.) 

With the above considerations, the equation (\ref{eq3}) for the 
magnetic permeability 
takes the form
\begin{equation}
\label{eq6}
\mu =\mu_0[1-\frac{{F}'\omega^2}{\omega^2-\omega_{LC}^2+i\omega \gamma}],
\end{equation}
with 
\begin{equation}
\label{eq7}
{F}'=F\frac{L}{L+L_e }, \, F=\frac{ltw}{V_{uc}}=
\frac{\mbox{inter-pair volume}}{\mbox{unit cell volume}}.
\end{equation}

Using the above formulas and their behavior going to  
small length scales we will show in the following that one can reproduce and 
explain all the high frequency magnetic response of artificial magnetic 
structures.

For that, it is important to notice that by scaling-down the structures 
uniformly, i.e. all the lengths scale proportionally to a basic length $a$ 
($=a_{\bf k}$ here), both the capacitance and the magnetic inductance scale 
proportionally to $a$, while the kinetic inductance and the resistance scale 
proportionally to $1/a$, i.e.
\begin{equation*}
C=\varepsilon_0 \varepsilon_b \frac{wl}{t} \sim a, \,
L=\mu_0 \frac{tl}{w} \sim a, \,  
\end{equation*}
\begin{equation}
\label{eq8}
R=\frac{\gamma_m}{\omega_p^2 \varepsilon_0}\frac{l}{wt_m}\sim \frac{1}{a}, 
\, {\rm and} \, L_e =\frac{l}{wt_m}\frac{1}{\omega_p^2 
\varepsilon_0}\sim\frac{1}{a}.
\end{equation}

The above formulas shows the increasingly pronounced role that kinetic 
inductance (also the resistance) plays in the smaller scales. 
Specifically, one can see that the 
ratio $L/L_{e}$ is of the order of $40 tt_{m}/\lambda_{p}^{2}$, with 
the typical 
value of $\lambda_{p}$ ($\lambda_p=2\pi c/\omega_p)$ being around 85 nm for
Al and 130 for Ag\cite{Boltasseva09}. 
Thus, for $\sqrt {tt_m }$ smaller than 100 nm the kinetic inductance, 
$L_{e}$, becomes appreciable and may dominate as the length scale becomes 
smaller and smaller.

\subsection{Magnetic resonance frequency}
Taking into account the expression 
for the magnetic resonance frequency 
in Eq. (\ref{eq3b}), in combination with Eqs. (\ref{eq8}), it 
can be shown that the magnetic resonance frequency of a slab-pair system 
 has the following scale dependence:

\begin{equation}
\label{eq9}
\omega_{LC}=\frac{1}{\sqrt{(L+L_e )C}}\propto \frac{1}{\sqrt{A_1a^2+A_2}},
\end{equation}
with $A_1$, $A_2$ constants (depending on the geometrical characteristics
of the structure). Eq. (\ref{eq9})  shows 
that as the slabs length-scale becomes smaller 
the magnetic 
resonance frequency does not continuously increase, but beyond a length-scale 
it saturates to a constant value. 
The saturation of the magnetic resonance frequency
is  exactly what is observed in  Fig.~2, and is exclusively
due to the existence of kinetic inductance (note that the resistance,
$R$, representing the ohmic losses, does not appear in the
above formula (\ref{eq9})). This inductance 
originates from the electrons inertia and
 represents the ``difficulty'' of electrons to follow high frequency motions,
 i.e. their difficulty to respond to high frequency fields.

Note that without the consideration of $L_e$ the second term in the 
square root of Eq. (\ref{eq9}) 
would not exist, leading to a linear dependence $\omega \sim 1/a$ (as occurs 
in  microwaves and larger scales) making unable to explain the 
saturation behavior observed in Fig. 2.
Note also that the kinetic inductance $L_e$ does not influence the 
response of the structure 
only at the magnetic resonance regime but at all high-frequency regimes,
including permittivity resonances.

Substituting Eqs. (\ref{eq8}) in Eq. (\ref{eq9}) 
and defining normalized (dimensionless) geometrical parameters, 
i.e. $l^{\prime}=l/a$, 
$t^{\prime}=t/a$, $t_m^{\prime}=t_m/a$,
one can obtain the dependence of $\omega_{LC}$
from the geometrical characteristics of the structure, as well as an
expression
for the
saturation value, $\omega_{LC}^{{\rm sat}}$:


\begin{equation}
\label{eq10}
\omega_{LC} =\frac{1}{\sqrt{\frac{\varepsilon_b {l^{\prime}}^2}{c^2} a^2+
\frac{{\varepsilon_b l^{\prime}}^2}{t^{\prime} t^{\prime}_m \omega_p^2}} }
\mathrel{\mathop{\kern0pt\longrightarrow}\limits_{a \to 0}} \omega_p 
\frac{\sqrt{t^{\prime}t^{\prime}_m} }{l^{\prime}\sqrt{\varepsilon_b}}=
\omega_{LC}^{{\rm sat}}.
\end{equation}

A correction to the above formula (\ref{eq10}) can be obtained if one takes 
into account also the potential energy of the electrons inside the 
metal slabs, 
through an equivalent capacitance\cite{Tretyakov07}, 
$C_e =\varepsilon_0 wt_m /l$ 
(the capacitance of the capacitor formed inside the metal), added to the 
slabs capacitance, $C$. In this case the saturation value for the magnetic 
resonance frequency is found as

\begin{equation}
\label{eq11}
\omega_{LC}^{\mbox{sat}}=\frac{\omega_p}{\sqrt 
{\frac{\varepsilon_b l^{\prime 2}}{t^{\prime}t^{\prime}_m }+1} },
\end{equation}
showing that the absolute upper limit for the saturation frequency is not 
arbitrary high but it is restricted by the plasma frequency of the bulk metal.

Since in the following we will consider systems with both 
$l^{\prime}/t^{\prime}$ and 
$l^{\prime}/t^{\prime}_m $ larger than unity, where the simplified equation 
(\ref{eq10}) is still 
valid, we will omit the electron potential energy in the following 
discussion, keeping into account though that this energy/capacitance sets a 
finite upper limit for the saturation value of the magnetic resonance 
frequency, which is the plasma frequency of the bulk metal.

\subsection{Magnetic permeability resonance}

As had been shown in Section II,
by reducing the length-scale of the artificial magnetic structures 
the magnetic permeability resonance becomes more and more weak, unable to 
lead to negative $\mu$ values beyond a length scale. This weakening is 
revealed in both the real part of  $\mu$ (where smaller absolute values of the 
Max[Re($\mu$)] and Min[Re($\mu$)] are observed) and the imaginary part 
(where smaller Max[Im($\mu$)], at $\omega =\omega_{LC}$, is 
observed) - see Fig.~3.

A detailed examination of Eq. (\ref{eq6}) reveals that the strength
of the resonance  is determined from both factors $\gamma$ and $F^{\prime}$.
On the other hand, the width of the negative permeability
regime, $\Delta\omega$, 
seems to be much more sensitive to the factor $F^{\prime}$, and almost
unaffected from $\gamma$ ($\gamma$ has only small influence on the lower
limit of the negative permeability band, while $F^{\prime}$ strongly affects
the upper limit of this band).  Note that in the absence of losses, i.e. 
$\gamma=0$, the upper limit of the
negative permeability band is $\omega_{LC}/\sqrt{1-F^{\prime}}$; the lower
limit is simply $\omega_{LC}$, i.e.
\begin{equation}
\label{Dw}
\Delta\omega=\omega_{LC}(\frac{1}{\sqrt{1-F^{\prime}}}-1).
\end{equation}

Using Eqs. (\ref{eq7}) and (\ref{eq3b}) in combination with with 
Eqs. (\ref{eq8}), one can
 obtain the scaling dependence of both $F^{\prime}$ and $\gamma$, as 

\begin{equation}
\label{eq12}
F^{\prime}=F\frac{L}{L+L_e}\propto B F a^2  
\end{equation}
($B$ constant) and
\begin{equation}
\label{eq13}
\gamma=\frac{R}{L+L_e} \propto \frac{1}{D_1 a^2+D_2}
\end{equation}
($D_1$, $D_2$ constants).
From Eq. (\ref{eq12}) one can derive two important conclusions:
(a)
The factor $F^{\prime}$,
which mainly determines the frequency width of the negative permeability 
regime, is independent of the resistance, $R$, thus independent of any loss
mechanisms.
(b)
With the consideration of  the kinetic inductance the factor $F^{\prime}$ from
scale
independent (if $L_e$ is negligible) becomes scale 
dependent and tends to zero as the size of the structure becomes 
smaller and smaller ($a\rightarrow 0$). This means that even in 
the absence of ohmic losses, it would be impossible to get
negative permeability values of band-width substantially larger
than $(a^2/\lambda_p^2)\omega_{LC}$ in arbitrarily small length scales.

The scaling behavior of $F^{\prime}$ also implies 
that the relative bandwidth of the negative
permeability regime, while it is maintained almost constant before the
staring of the saturation (see Fig. 4), it becomes smaller 
and smaller deeper in the saturation regime, indicating that working before
the saturation regime favors the widest bandwidth of each specific structure. 

The geometrical dependence of the factor $F^{\prime}$ for the slab-pair
design can be easily obtained
by substituting Eqs. (\ref{eq8}) to Eq. (\ref{eq12}):
\begin{equation}
\label{eq14}
F^{\prime}=F\frac{1}{1+c^2/(\omega_p^2 t_m t)}=
\frac{F}{1+\lambda_p^2/[(2\pi)^2 t t_m]}.
\end{equation}

Concerning the loss factor $\gamma$, from Eq. (\ref{eq13}) one can see that 
 $\gamma$ increases
as the length-scale decreases, justifying the higher losses in the smaller
length scales. This  increase though does not continue up to the
smallest scales, but beyond a length-scale it
approaches a saturation value. The geometrical 
dependence of $\gamma$ for the slab-pair design is obtained analogously 
with that of $F^{\prime}$, as:

\begin{equation}
\label{eq15}
\gamma=\frac{\gamma_m}{1+(\omega_p^2/c^2) t_m t}
=\frac{\gamma_m}{1+(2\pi)^2 t_m t/\lambda_p^2}.
\end{equation}
Eq. (\ref{eq15}) shows that the saturation value of $\gamma$ is
the collision frequency of the bulk metal as considered in the
free-electron (Drude) description of the metal\cite{comment}.
The saturation regime for  $\gamma$ is approached simultaneously with
the magnetic resonance frequency saturation, showing that deep 
in the saturation
regime the weakening of the resonance is not mainly the result of the 
ohmic losses but it is rather the result of the kinetic inductance (affecting
through the factor $F^{\prime}$).

Finally, it is important to point out that both $\gamma$ and $F^{\prime}$
depend not only on the ``quality'' (i.e. plasma frequency and damping factor)
of the metal used for the fabrication of metamaterials but also on the
geometrical parameters of the structures (see Eqs. (\ref{eq14}) 
and (\ref{eq15})). This reveals the  possibility to modify these factors,
and thus to enhance the performance
of the  high frequency metamaterials, by modifying the geometry.

\section{Optimized designs}

From the analytical formulas and the discussion of the previous section
it becomes clear that for achievement of optimized high frequency
(e.g. optical) magnetic metamaterials 
one should require:
(a) Highest possible saturation value for the magnetic resonance frequency;
(b) larger possible parameter $F^{\prime}$, determining the strength of the
magnetic resonance and the width of the negative permeability regime;
(c) smallest possible loss factor $\gamma$. 

The general requirements for meeting the above conditions can be easily
concluded based on Eqs. (\ref{eq9}),  (\ref{eq12}) and (\ref{eq13}). They 
demand:
Structures of small capacitance, $C$; structures of small kinetic 
inductance $L_e$, compared to magnetic inductance, $L$; structures of low
resistance, $R$. On the other hand, 
the role of the magnetic field inductance is more puzzling:
while low inductance facilitates the  achievement of high magnetic resonance 
frequency, it results to ``lower-quality'' resonance, i.e. weaker 
resonance and higher losses (compared to a higher inductance structure of the
same length scale). Thus, the inductance optimization 
should be based on the specific requirements 
for the designed metamaterial.

To translate the above optimization conditions to specific geometrical
and material conditions for the slab-pair-based 
systems we can use Eqs. (\ref{eq10}), (\ref{eq14}) and (\ref{eq15}). 
From these equations can be concluded that optimized slab-pair-based systems
are favored from
structures of thick metal (high $t_m$) and thick separation layer between
the slabs of the pair (i.e. high $t$ - not as high though as to cancel the 
interaction between the slabs); also structures of wide slabs (i.e. 
large width $w$ - to maximize the
structure volume fraction $F$ appearing in Eq. (\ref{eq14})).
Moreover, high quality optical metamaterials demand metals of the highest
possible plasma frequency and the lowest possible collision frequency.

The role of the slab-length is not one-way:
while shorter slabs (compared to the corresponding unit cell side) 
facilitate high attainable magnetic resonance frequencies
they lead to narrower negative permeability regime (due to reduced $F$ in
Eq. (\ref{eq14})),  and vice versa.

Finally, we should emphasize here that for enhanced  metamaterial
performance one should target operation below the saturation regime, 
as to ensure large negative permeability width and lower losses.

\subsection{The fishnet design}

As was mentioned in the previous sections and can be easily concluded from 
Eq. (\ref{eq9}), structures with reduced  inductance 
lead to higher 
magnetic resonance frequencies. A structure based on slab-pairs which offers 
a considerable reduction in the inductance is the fishnet design
- see  Fig. 1(d). Studies of this design in microwaves\cite{Kafesaki07} 
have revealed that at the magnetic resonance loop currents exist
not only at the slab pair but 
also in the necks' part of the metallic element.
This neck contribution can be taken into account in 
an effective LC circuit description of the structure by considering an 
additional inductance, due to the necks, which is in parallel 
with the inductance of the slabs, resulting to  a reduced total 
inductance, $L_{\rm fishnet}$, with
\begin{equation}
\frac{1}{L_{\rm fishnet}}= \frac{1}{L_{\rm slabs}}+
\frac{1}{L_{\rm necks}}
\end{equation}
($L_{\rm slabs}$ and $L_{\rm necks}$ represent the inductance of the
slabs and the necks part respectively; note that the $L_{\rm slabs}$
here it is not exactly the same as that for the slab-pair-only system,
 due to
the difference in the charge and current distribution
between only-slab-pair systems and fishnet.) 
This reduced inductance
leads to a magnetic resonance frequency
\begin{equation}
\label{eq-final}
\omega_{LC, \, {\rm fishnet}}^{\rm  2}=\frac{1}{L_{\rm fishnet}C} \approx
\omega_{LC, \, {\rm slabs}}^{\rm 2}(1+\frac{L_{\rm slabs}}{L_{\rm necks}}), 
\end{equation}
i.e. higher than that of only the slabs.

By reducing the size of the structure down to submicron and nanometer
scale, the inductance of both the slabs and the necks part gets a  
contribution from the kinetic (electrons) inductance.
This contribution though does not modify the general relation
(\ref{eq-final}), indicating the  higher magnetic resonance frequency
of the fishnet design than that of the component wide-slab-pair system
even in the saturation regime.

This higher attainable magnetic resonance frequency makes the fishnet
design the preferential one for the achievement of optical metamaterials,
something that has been already revealed from many existing
 experimental and theoretical works
concerning optical metamaterials\cite{Soukoulis07a,Shalaev07}.

The relation (\ref{eq-final}), combined with Eqs. (\ref{eq8}), can easily
lead to a relation for the geometrical dependence of the magnetic resonance
frequency for the fishnet, i.e.
\begin{equation}
\label{eq-final2}
\omega_{LC, \, {\rm fishnet}}^{\rm  2}=
\omega_{LC, \, {\rm slabs}}^{\rm 2}(1+\frac{w_n}{w}
\frac{l}{a_{\bf E}-l}),  
\end{equation}
and thus for the saturation value of this magnetic resonance frequency.
In Eq.  (\ref{eq-final2})
$w_n$ and $w$ is the width of the neck and slab
parts respectively (along {\bf H}-direction) 
and 
$a_{\bf E}$ is the lattice constant along the ${\bf E}$
direction (see Fig. 1). 
Eq. (\ref{eq-final2}) suggests that for the achievement of high magnetic
resonance frequency saturation 
values for fishnet, apart from the conditions for the optimization
of the slab-pair components one should pursue also wide neck parts; moreover,
 the wider
the neck parts the higher the saturation value of the fishnet
magnetic resonance frequency is.

Concerning the magnetic permeability expression for the fishnet, here
the situation is more complicated compared to only-slabs systems, 
due to the more complicated current 
picture at the magnetic resonance\cite{Kafesaki07}.
Following the observations and conclusions of Ref. \cite{Kafesaki07}, 
according to which the fishnet unit cell can be approximated  
with an RLC circuit with inductance of slabs and inductance
of necks connected in parallel, the magnetic permeability 
for the fishnet can be calculated 
following the same steps as the ones presented in the previous section 
 for the slab case, 
with modifications
in the incident flux,  the magnetic moment and the total inductance
per unit cell:

Here the 
incident flux, $\phi$, can be written as
$\phi=\mu_0 a_{\bf E} t H_0 {\rm e}^{-i \omega t}$, and the modified
magnetic moment as

\begin{equation}
\label{eq-m-fishnet}
m= I_{\rm slabs}lt-I_{\rm necks}(a_E-l)t =I L_{\rm fishnet} 
[\frac{lt}{L_{\rm slabs}}-\frac{(a_{\bf E}-l)t}{L_{\rm necks}}],
\end{equation}
where $I=I_{\rm slabs}+I_{\rm necks}$ and 
$I L_{\rm fishnet}=I_{\rm slabs}L_{\rm slabs}=
I_{\rm necks} L_{\rm necks}$.

The inductances 
$L_{\rm slabs}$ and $L_{\rm necks}$ include both the magnetic field inductance
and the electrons inductance for slabs and necks.

From Eq. (\ref{eq-m-fishnet}) one can see already 
that the presence of the necks weakens
the magnetic response of the structure, since the neck 
contribution in the magnetic moment opposes that of the slabs.

With the above considerations, the magnetic permeability
 for the
fishnet design can be expressed as

\begin{equation}
\label{eq23}
\mu(\omega)=\mu_0 [1-\frac{(1/V_{uc})
\mu_0 a_E t\omega^2 
(lt/L_{\rm slabs}-(a_E-l)t/L_{\rm necks})
}{\omega^2
-1/L_{\rm fishnet}C+i\omega R/L_{\rm fishnet}}].
\end{equation}

Comparing the above equation with Eq. (\ref{eq3}), one can observe
that the factor multiplying the $\omega^2$ in the numerator,
which is the main factor determining the spectral width
of the negative permeability regime, becomes smaller than that of the 
slab only
case ($F^{\prime}$). This can explain the reduced 
(compared to only slabs) spectral
width of the fishnet negative $\mu$ regime observed in Fig. 4.

\section{Obtaining quantitative results}

In the previous sections we presented a simple RLC circuit model 
for obtaining
qualitative results and optimizing the slab-pair-based metamaterial
structures. 
It is important to note though that the relations (3)-(4) 
do not only have a qualitative 
power, but they can be used
also to obtain quantitative results, if plugged with more accurate expressions
for the capacitances and inductances involved.

Here we will demonstrate this quantitative power of our relations in the
case of the wide-slab-pair system. For wide-slab-pair systems ($w=a_{\bf H}$)
and examining the fields and currents at the magnetic resonance frequency,
one can use for the total capacitance of the system the capacitance 
of two identical  parallel plate capacitors  connected in-series, each of 
capacitance $C= \epsilon_0\epsilon_b(w/t)(l/2.5)$,
and for the electrons inductance the same relation as in Eq. (9), multiplied
by 2 (to take into account both slabs of the pair).
 
The calculated magnetic resonance frequency obtained by using these relations
is shown with the dotted-dashed line in Fig. 6, 
and as can be seen there
it is in excellent agreement with the  magnetic resonance frequency
obtained through realistic calculations.

Using the same equations 
for the case of narrow slabs, one can see that  they
fail to describe accurately the scaling of the magnetic resonance frequency. 
In fact these equations predict no dependence of $\omega_m$  on 
the width of the slabs, unlike the
observed in Fig. 2 result, which shows lower magnetic resonance frequency for
the narrower slabs.

To explain this discrepancy, we should point out that as the 
slabs become narrower
Eqs. (9) for the capacitance and the magnetic field inductance
become less and less accurate, and in the limit of $w<t$ they should be 
replaced
by the corresponding expressions for parallel wire systems, i.e.
$L=\mu_0 l \ln(t/w)/\pi$, and   $C =\epsilon_b \epsilon_0 \pi(l/2.5)/\ln(t/w)$.
The last relations for $L$ and $C$ lead to lower magnetic resonance frequency
saturation values compared to the ones predicted by the corresponding
Eqs. (9), in agreement with the numerical results presented in
Fig. 2.


\section{Conclusions}
In this paper we examine the magnetic response of
resonant magnetic  structures based on the slab-pair design, 
as the structures are scaled down from mm to nm scale.
This response is examined using detailed numerical simulations
for obtaining the frequency dependence of the magnetic permeability 
(including its resonant behavior) through reflection
and tranmission data of realistic structures.
It is observed, as expected, 
that the magnetic resonance frequency of the structures,
while it scales inversely proportional to the structure length scale in
the mm scale,  it saturates to a constant 
value in the nano regime. This behavior depends on
the design and it is independent of any ohmic losses in the structure. 
Among our
designs, higher
saturation value is observed for the fishnet design.

The permeability resonance becomes
 weaker and weaker as we go  deeper into sub-$\mu$m scale, 
and ultimately 
it does not
reach negative values. The relative spectral 
width of the negative permeability regime
which in larger scales is almost scale independent, in sub-$\mu$m scales it
becomes smaller and smaller, approaching zero, and it has very slight 
dependence on the ohmic losses, while it shows strong dependence
on the design, being quite narrow for the fishnet design. 

All the above results are explained through a simple RLC circuit model;
in the inductance, $L$, the current-connected kinetic energy of the
electrons is taken into account besides the magnetic
field energy.
This model is capable to determine optimization conditions for 
our structures, 
as to attain high magnetic resonance frequencies maintaining strong
resonant response with a negative permeability region as wide as possible.
The model explains also the superior performance of the fishnet design
regarding high magnetic resonance frequency and its reduced performance
regarding the width of the negative permeability region.
Finally, we show that our simple RLC circuit model is capable not only
to predict qualitatively the behavior of our structures, but also to give
quantitative results if it is accompanied by accurate formulas for
the capacitance and inductance of the systems.

\section{Acknowledgments}

Authors would like to acknowledge financial support by the
European Union FP7 projects PHOME, ENSEMBLE, ECONAM, NIMNIL, and
the COST Actions MP0702, MP0803, by the European Office of Aerospace Research
and Development (under project FENIM), and by the US Department of 
Energy (contract no. DE-AC02-07-CH11358).


\begin{thebibliography}{99}
\bibitem{comment}
Usually in simulations  concerning realistic structures the
factor $\gamma$ employed is slightly larger than that of the bulk
metal, to account for the surface scattering effects.
\end{thebibliography}
\end{document}